\documentclass[aps,preprint,showpacs,superscriptaddress]{revtex4-1}

\usepackage{amssymb}
\usepackage{mathrsfs}
\usepackage{lipsum}
\usepackage{graphicx}
\usepackage[caption=false]{subfig}
\usepackage{mathtools}
\usepackage{epstopdf}
\usepackage{xcolor}
\usepackage[colorlinks,urlcolor=blue]{hyperref}
\DeclareGraphicsExtensions{.pdf,.jpg,.png,.eps}
\usepackage[toc,page,title,titletoc,header]{appendix}
\begin{document}
	
	\title{Quantum process tomography with Heisenberg scaling based on Gaussian state and binary detection}
	
	\author{Jian-Dong Zhang}
	\author{Zi-Jing Zhang}
	\email{Corresponding author: zhangzijing@hit.edu.cn}
	\author{Long-Zhu Cen}
	\author{Yuan Zhao}
	\email{Corresponding author: zhaoyuan@hit.edu.cn}
	\affiliation{Department of Physics, Harbin Institute of Technology, Harbin, 150001, China}

\begin{abstract}
We propose a quantum process tomography scheme that utilizes two-mode squeezed vacuum to realize the parameter estimation with Heisenberg scaling.
The objective is to estimate a rotating angle of polarization and parity detection is used as the detection strategy.
With the help of symplectic matrix theory, we discuss the estimation visibility and sensitivity of output signal in lossless situation, the quantum Fisher information is also given via calculation. 
Finally, the impacts of two realistic factors on both visibility and sensitivity are also considered, including photon loss during the input generation, and photon loss along with thermal noise during the output detection.
\end{abstract}


\maketitle
 
\section{Introduction}

Quantum process tomography (QPT) \cite{PhysRevLett.93.080502, PhysRevA.64.012314, PhysRevLett.100.190403, PhysRevLett.97.220407, PhysRevA.87.062119, PhysRevA.82.042307, PhysRevLett.104.123601, PhysRevLett.105.053201}$\---$which is a significant application of high-precision parameter estimation$\---$has the purpose of accurately acquiring the information about estimated parameter. 
An everlasting objective of QPT is to achieve the sensitivities that are beyond the limit of classical parameter estimation. 
Generally, the estimation sensitivity is positively correlated with the input mean photon number $N$ used for estimation process. 
To sensitivities with $N^{-1/2}$ and $N^{-1}$ there correspond to shot-noise limit (SNL) and Heisenberg limit (HL), respectively. 
Hence, the probe states with large mean photon number have a greater sensitivity compared with small one under the appropriate strategies. 
Since the large photon number for coherent state can be easily prepared, it is most commonly used in QPT although its sensitivity is limited by the SNL. 
This is not an annoyance in the case of limitless resources or the objects that can withstand high-strength illumination. 
However, in some scenarios where only weak light can be applied \cite{Taylor2013Biological, wolfgramm2013entanglement}, the estimation sensitivity of coherent state is inferior to that of exotic quantum states, Schr\"odinger cat state \cite{ZHANG201792}, N00N state \cite{N00N}, and squeezed vacuum state \cite{PhysRevD.23.1693}, to name a few. 
Therefore, it is of great practical significance to break the SNL and to achieve preferable sensitivity via utilizing the same photons.

Recently, a great deal of schemes based on exotic quantum states are proposed for striving to realize super-sensitivity QPT \cite{cen2017state, Zhou15, PhysRevX.4.041025, PhysRevLett.112.103604, Afek879}. 
The two most excellent states among these schemes are two non-Gaussian ones: N00N state \cite{N00N} and Holland-Burnett (HB) states \cite{PhysRevLett.71.1355}.
They are reported to be sensitive to the Heisenberg scaling, unfortunately, these non-Gaussian schemes have their own defects. 
The state's fidelity of N00N with exact $N$ value is not ideal, for this process involves complex post-selection. 
Meanwhile, the probability of successful selection generally decreases with the increase of $N$ value.
On the other hand, biphoton or four-photon HB state can be generated by using type-I or type-II entangled source \cite{Zhou15}, whereas the preparation of HB state which consists more photons without the post-selection is still a conundrum that needs to be solved.
In contrast, the preparations of Gaussian states are relatively easy.

For improving such a predicament, in this paper, we propose a QPT scheme for estimating polarized rotating angle based on a Gaussian state: two-mode squeezed vacuum (TMSV) \cite{anisimov2010quantum, zhang2017effects}. 
The system sensitivity with parity detection is discussed and ultimate limit is given by calculating the quantum Fisher information (QFI) \cite{Gibilisco2007}. 
We additionally compare the sensitivity of our scheme and that of previous schemes, N00N state and HB state, in the case of employing the identical photon number.
Finally, we also analyze the impacts of the realistic factors$\---$photon loss in imperfect state generation, and photon loss accompanied by thermal noise for imperfect detector$\---$on estimation results.

\section{Fundamental principle}

TMSV state, also known as an Einstein-Podolski-Rosen (EPR) state, is an entangled state containing strong entanglement between the two modes. 
It has a superposition of twin Fock state, $\left| {n,n} \right\rangle $, representation as $\left| {{\psi _{\rm{in}}}} \right\rangle  = \sum\nolimits_{n = 0}^\infty  {\sqrt {\left( {1 - t} \right){t^n}} {{\left| {n,n} \right\rangle }}}$ in terms of different weights. 
Where ${\left| {n,n} \right\rangle } \equiv {\left| n \right\rangle }\otimes{\left| n \right\rangle }$, $t = {N \mathord{\left/
		{\vphantom {N {\left( {N + 2} \right)}}} \right.
		\kern-\nulldelimiterspace} {\left( {N + 2} \right)}}$, and $N$ is mean photon number of the TMSV state. 
One can find that only the paired state (HB state) occurs in TMSV state, moreover, it can be proved that the degree of entanglement increases with increasing $N$ by calculating the von Neumann entropy of reduced density matrix.

\begin{figure*}[htbp]
\centering
\includegraphics[width=\textwidth]{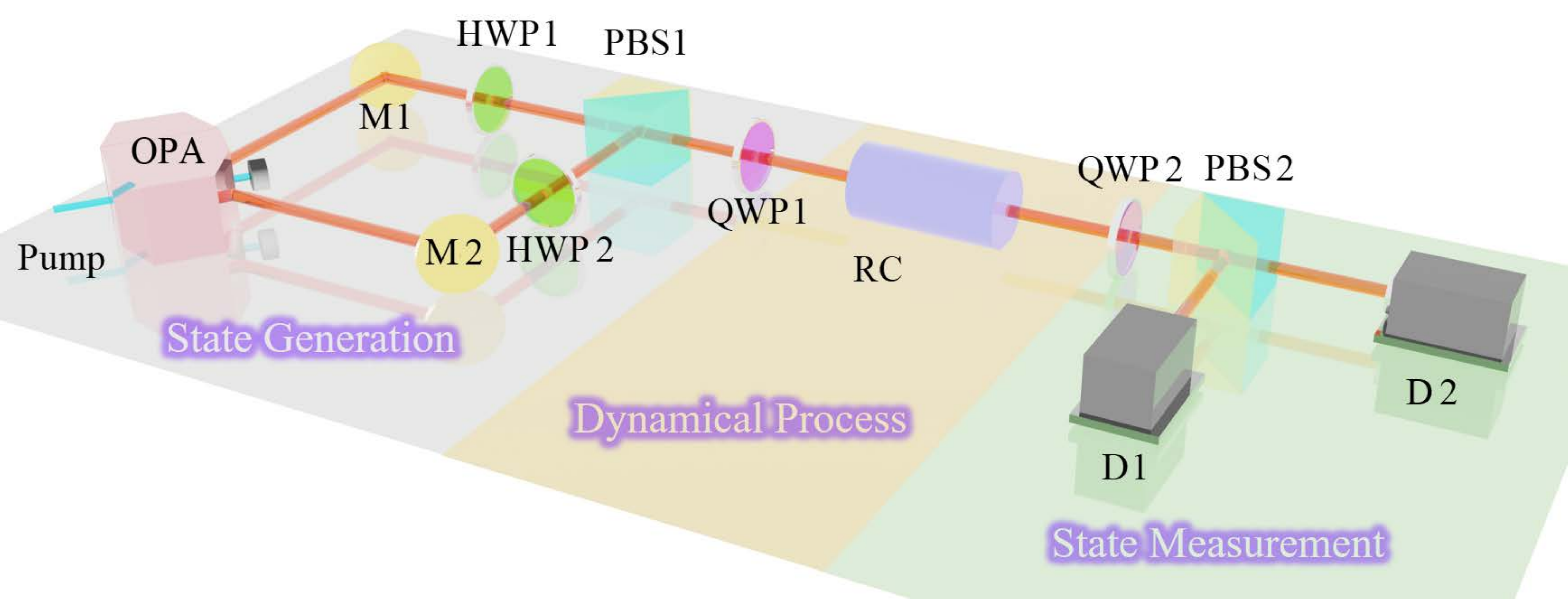}
\caption{Schematic of QPT scheme for rotating angle estimation of polarization. The TMSV state is prepared by an OPA, and its polarization mode is modulated by HWPs and QWPs. RC plays the role of generating a rotating angle. The output signal is detected by single detector and parity detection is performed. OPA, optical parametric amplifier; M, mirror; HWP, half wave plate; QWP, quarter wave plate; RC, rotating crystal; PBS, polarizing beam splitter; D, detector.}
\label{system}
\end{figure*}

The mode of the TMSV state in this paper refers to the polarization mode, and more explicitly, the fault probe state is $\left| {{\psi _{\rm{in}}}} \right\rangle  =  \sum\nolimits_{n = 0}^\infty   {\sqrt {\left( {1 - t} \right){t^n}} } {\left| {n,n} \right\rangle _{HV}}$.
The schematic of process tomography scheme with TMSV state is shown in Fig. \ref{system}.
The input state is generated by an optical parametric amplifier (OPA), and the two modes are modulated into horizontal and vertical polarization by half wave plates (HWPs), respectively.
Then the two different polarized modes are coupled through the first polarizing beam splitter (PBS1) and output from the same port. 
Subsequently, the state passes through the first quarter wave plate (QWP1), linearly polarized modes are transformed into circularly polarized ones. 
At this time, the state can be written as $\left| {{\psi ^*_{\rm{in}}}} \right\rangle  =  \sum\nolimits_{n = 0}^\infty   {\sqrt {\left( {1 - t} \right){t^n}} } {\left| {n,n} \right\rangle _{LR}}$, and then it experiences 
a unitary rotating channel with a rotating angle $\theta$ for polarized azimuth, which can be simulated by a HWP or a Faraday rotation crystal \cite{PhysRevLett.112.153601}. 
After rotating channel, the state is remodulated into linearly polarized modes by the second quarter wave plate (QWP2).
Finally, the state goes though the second polarization beam splitter (PBS2) and parity detection strategy is performed for outcome.

Parity detection is primitively proposed by Bollinger $et$ $al$. for enhanced frequency measurement with an entangled state of trapped ions \cite{bollinger1996}. 
Later, Gerry and Campos applied it to quantum metrology for achieving Heisenberg scaling estimation sensitivity \cite{gerry2000heisenberg, gerry2005quantum}. 
In this strategy, the parameter information is obtained by binarizing the photon number$\---$in terms of the parity$\---$in either of the two output ports, i.e., it is more concerned with the parity of photon number at the output rather than exact number of photons. 
The single measurement result is denoted as +1 if the photon number is even and $-1$ if odd. 
Take the horizontal output port as an example (port $B$), the parity operator can be written as $\left\langle {\hat \Pi_B } \right\rangle  = \exp \left( {i\pi \hat a_H^\dag {{\hat a}_H}} \right)$.

\section{Lossless situation}
In this section, we offer a calculation method using the theory of symplectic matrix, the whole process is analyzed in phase space and the details can be found in Ref. \cite{RevModPhys.84.621}.
The role of QWPs in our scheme is to switch the polarized modes between linear polarization and circular one. 
The relationship between the creation operator of linear polarization and that of circular one (before and after QWP) is described as follows\cite{PhysRevLett.117.103601}:
\begin{eqnarray}
\hat a_R^\dag  = \frac{1}{{\sqrt 2 }}\left( {\hat a_H^\dag  + \hat a_V^\dag } \right), \\ 
\hat a_L^\dag  = \frac{1}{{\sqrt 2 }}\left( {\hat a_H^\dag  - \hat a_V^\dag } \right). 
\end{eqnarray}
Further, we can obtain the matrix forms of QWPs in accordance with theirs effect in polarized mode conversion,
\begin{equation}
{\mathbf{S}_{\rm{QWP1}}} = {\mathbf{S}_{\rm{QWP2}}} = \frac{1}{{\sqrt 2 }}\left( {\begin{array}{*{20}{c}}
	1 & 0 & 1 & 0  \\
	0 & 1 & 0 & 1  \\
	1 & 0 & { - 1} & 0  \\
	0 & 1 & 0 & { - 1}  \\
	\end{array}} \right).
\end{equation}
For a rotating crystal, the left- and right-handed polarization modes are respectively introduced the phase of $\exp\left( i\theta\right) $ and $\exp\left( -i\theta\right) $ \cite{d2013photonic} as left- and right-handed polarized states are eigenstates of rotating operation.
So we can write the matrix of rotating crystal as
\begin{equation}
{\mathbf{S}_{\rm{RC}}} = \left( {\begin{array}{*{20}{c}}
	{\cos \theta } & { - \sin \theta } & 0 & 0  \\
	{\sin \theta } & {\cos \theta } & 0 & 0  \\
	0 & 0 & {\cos \theta } & {\sin \theta }  \\
	0 & 0 & { - \sin \theta } & {\cos \theta }  \\
	\end{array}} \right).
\end{equation}

In addition, the input mean and covariance matrix of TMSV state have the following matrix forms in phase space \cite{RevModPhys.84.621}
\begin{equation}
{{\bf{M}}_{\rm{in}}} = {\left( {\begin{array}{*{20}{c}}
		0 & 0 & 0 & 0  \\
		\end{array}} \right)^\top},
\end{equation}
\begin{equation}
{\mathbf{\Gamma} _{\rm{in}}} = \left( {\begin{array}{*{20}{c}}
	{N + 1} & 0 & {\sqrt {N\left( {N + 2} \right)} } & 0  \\
	0 & {N + 1} & 0 & { - \sqrt {N\left( {N + 2} \right)} }  \\
	{\sqrt {N\left( {N + 2} \right)} } & 0 & {N + 1} & 0  \\
	0 & { - \sqrt {N\left( {N + 2} \right)} } & 0 & {N + 1}  \\
	\end{array}} \right).
\end{equation}

By implementing the following transformations, 
\begin{align}
{{\bf{M}}_{{\mathop{\rm out}\nolimits} }} &= {\bf{S}}{{\bf{M}}_{\rm in}}, \\
{\Gamma _{{\mathop{\rm out}\nolimits} }} &= {\bf{S}}{\Gamma _{\rm in}}{{\bf{S}}^\top}, 
\end{align}
with $\mathbf S = {\mathbf S_{\rm{QWP2}}}{\mathbf S_{\rm{RC}}}{\mathbf S_{\rm{QWP1}}}$, we obtain the output mean and covariance matrix.
Then, we can calculate the expectation value of the parity operator \cite{NJP113025},
\begin{equation}
\left\langle {{{\hat \Pi }_{B}}} \right\rangle  = \frac{{\exp \left( { - \mathbf{M}_{\textrm{out}\left( {3,4} \right)}^\top \mathbf \Gamma _{\textrm{out}\left( {3,4} \right)}^{ - 1}  {\mathbf{M}_{\textrm{out}\left( {3,4} \right)}}} \right)}}{{\sqrt {\left| {{\mathbf\Gamma _{\textrm{out}\left( {3,4} \right)}}} \right|} }}=\frac{1}{{\sqrt {1 + N\left( {N + 2} \right){{\cos }^2}\left( {2\theta } \right)} }},
\end{equation}
where ${\mathbf \Gamma_{\textrm{out}\left( {3,4} \right)}}$ is the lower right submatrix of output covariance matrix and ${\mathbf{M}_{\textrm{out}\left( {3,4} \right)}}$ is the last two elements of output mean. Since the input mean is a full-zero matrix, the output ${\mathbf{M}_{\textrm{out}\left( {3,4} \right)}}$ only contains zero elements.

According to the definition of visibility \cite{N00N},
\begin{equation}
V = \frac{{{{\left\langle {{\hat \Pi_{B}}} \right\rangle }_{\max }} - {{\left\langle {{\hat \Pi_{B}}} \right\rangle }_{\min }}}}{{\left| {{{\left\langle {{\hat \Pi_{B}}} \right\rangle }_{\max }}} \right| + \left| {{{\left\langle {{\hat \Pi_{B}}} \right\rangle }_{\min }}} \right|}},
\label{visibility}
\end{equation}
we can calculate the visibility of our scheme, $V=N/\left( N+2\right)$.
For large $N$, the signal has an approximate 100\% visibility.

Then, we can obtain the classical Fisher information \cite{PhysRevA.62.012107}, 
\begin{equation}
{F_{\rm{c}}} = \frac{1}{{P_{\rm{e}}^2}}{\left( {\frac{{\partial {P_{\rm{e}}}}}{{\partial \theta }}} \right)^2} + \frac{1}{{P_{\rm{o}}^2}}{\left( {\frac{{\partial {P_{\rm{o}}}}}{{\partial \theta }}} \right)^2} = {\left[ {\frac{{2\sqrt {N\left( {N + 2} \right)} \sin \left( {2\theta } \right)}}{{1{\rm{ + }}N\left( {N + 2} \right){{\cos }^2}\left( {2\theta } \right)}}} \right]^2},
\end{equation}
with
\begin{eqnarray}
{P_{\rm{e}}} = \frac{1}{2}\left( {1 + \left\langle {{{\hat \Pi }_B}} \right\rangle } \right), \\ 
{P_{\rm{o}}} = \frac{1}{2}\left( {1 - \left\langle {{{\hat \Pi }_B}} \right\rangle } \right). 
\end{eqnarray}
By means of the relationship between the sensitivity and classical Fisher information, we get the optimal sensitivity,
\begin{equation}
\delta {\theta _{\min }} = {\left. {\frac{1}{{\sqrt {{F_{\rm c}}} }}} \right|_{\theta  = {\pi  \mathord{\left/
				{\vphantom {\pi  4}} \right.
				\kern-\nulldelimiterspace} 4}}} = \frac{1}{{2\sqrt {N\left( {N + 2} \right)} }}.
\end{equation}
Additionally, based upon the matrix calculation method for quantum Fisher information in Ref. \cite{PhysRevA.94.063840}, the quantum Cram\'er-Rao bound is supplied with
\begin{equation}
F_{\rm q} = \frac{1}{{2\sqrt {N\left( {N + 2} \right)} }}.
\end{equation}

One can see that the sensitivity of parity detection is saturated with quantum Cram\'er-Rao bound and is better than the Heisenberg limit ($1/2N$), this indicates that our strategy is optimal scheme and obtains sub-Heisenberg-limited sensitivity.
The above results suggest that our scheme not only bypasses the problem of low photon number in the input, but also achieves the sensitivity which is superior to the sensitivities of non-Gaussian schemes \cite{N00N, PhysRevLett.71.1355}.

\section{Realistic factors}
In the above section, we discussed the estimation signal and sensitivity under lossless condition, however, the impacts of realistic factors in practical detection are inevitable. 
In this section, we mainly study two realistic factors: imperfect input generation and output detection.

\subsection{Imperfect input generation}
For exotic quantum states such as TMSV, the imperfect input generation is widespread. 
A universal method of simulation analysis is to place two virtual beam splitters (VBSs) after the ideal input generated by OPA \cite{kacprowicz2010experimental, PhysRevA.80.013825}. 
The transmissivities of two VBSs are $T_1$ and $T_2$, and the photons entering the environment through VBSs reflection are regarded as loss.
For the need of photon number conservation, the dimensions of all matrices change from four-by-four to eight-by-eight as two environment ports are considered.
Two parts$\---$probe part and environment part$\---$make up the direct sum space.
The explicit matrix forms can be found in Appendix, and the transformation matrix $\bf S$ in Eqs. (7) and (8) is rewritten as $\mathbf S^{\rm R1} = {\mathbf S_{\rm{QWP2}}^{\rm R1}}{\mathbf S_{\rm{RC}}^{\rm R1}}{\mathbf S_{\rm{QWP1}}^{\rm R1}}{\mathbf S_{\rm{VBS}}^{\rm R1}}$ at this time. 

Using the calculation method in Eq. (9) and classical Fisher information, we can obtain the output signal and sensitivity,
\begin{equation}
{\left\langle {{{\hat \Pi }_B}} \right\rangle _{{\mathop{\rm R1}\nolimits}}} = \frac{1}{{\sqrt {{G_1}} }},
\end{equation}
\begin{equation}
\delta \theta  = \frac{{2\sqrt {1 - G_1^{ - 1}} }}{{\left| {{G_2}G_1^{ - {3 \mathord{\left/
						{\vphantom {3 2}} \right.
						\kern-\nulldelimiterspace} 2}}} \right|}},
\end{equation}
with
\begin{align}
\nonumber {G_1} =& {\left( {1 + N{T_1}} \right)^2}{\sin ^4}\theta  + {\left( {1 + N{T_2}} \right)^2}{\cos ^4}\theta + 2\left[ {1 + N\left( {{T_1} + {T_2} - N{T_1}{T_2}} \right)} \right]{\sin ^2}\theta {\cos ^2}\theta \\ 
&+ N{T_1}{T_2}\left[ {\cos \left( {4\theta } \right) - 1} \right], \\
{G_2} =& N\sin \left( {2\theta } \right)\left\{ {\left( {{T_2} - {T_1}} \right)\left[ {2 + N\left( {{T_1} + {T_2}} \right)} \right] + \left[ {8{T_1}{T_2} + N{{\left( {{T_1} + {T_2}} \right)}^2}} \right]\cos \left( {2\theta } \right)} \right\}.
\end{align}

\begin{figure}[htbp]
	\centering
	\includegraphics[width=8cm]{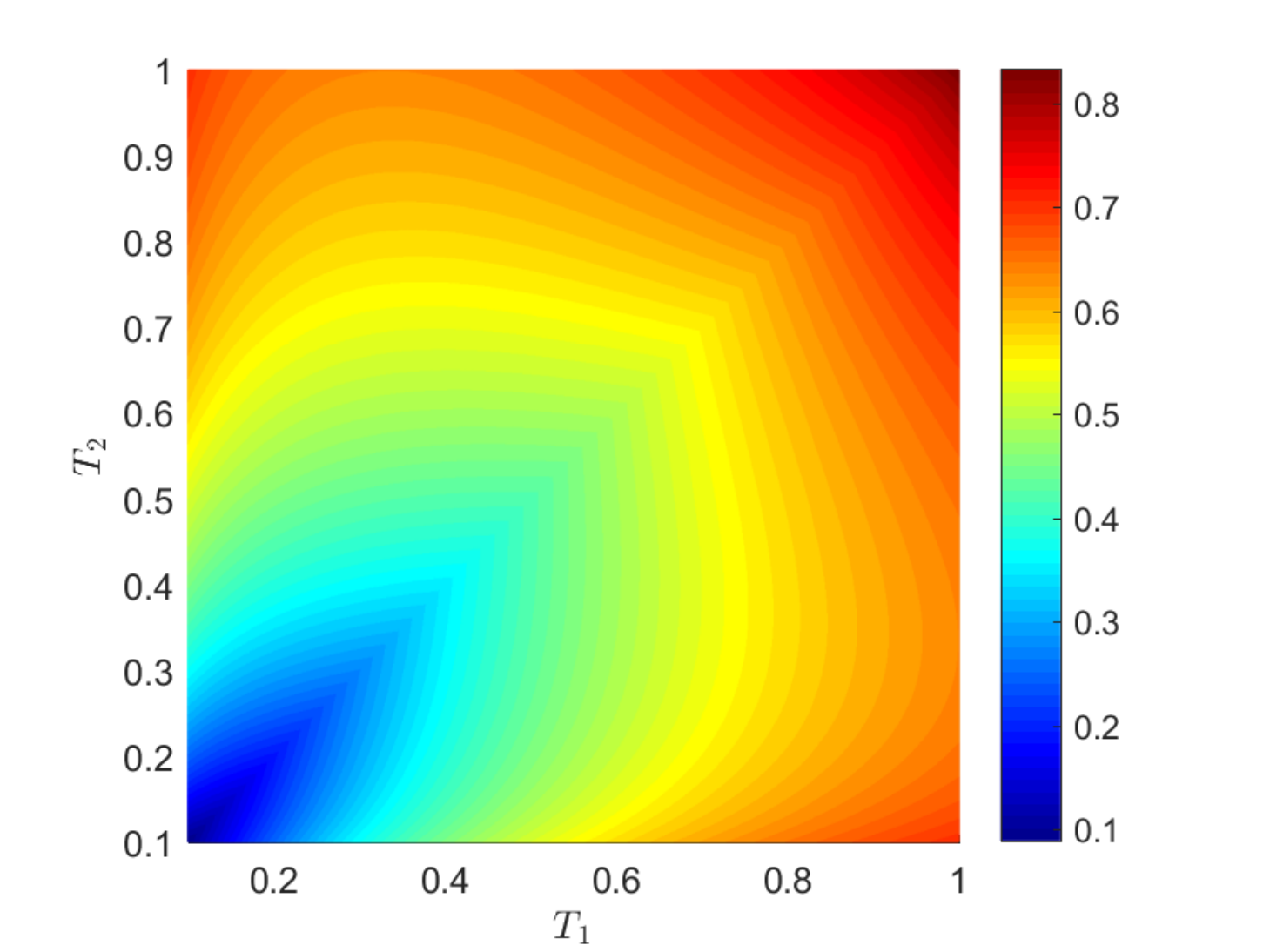}	
	\includegraphics[width=8cm]{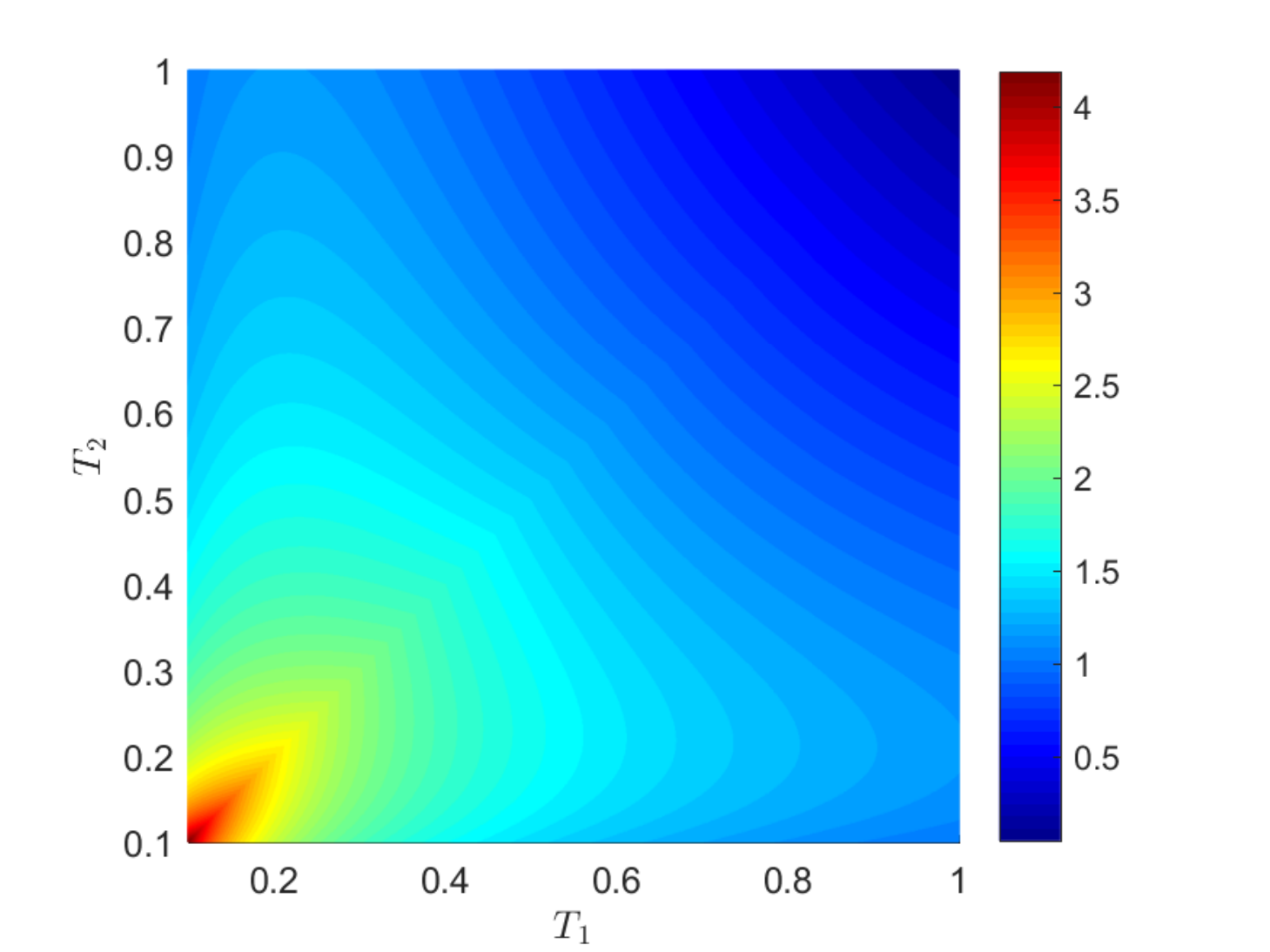}
	\caption{(a) Visibility with parity detection as a function of transmissivities, $T_1$ and $T_2$, with $N=10$. Both $T_1$ and $T_2$ range from 0.1 to 1. (b) Optimal sensitivity with parity detection as a function of transmissivities, $T_1$ and $T_2$, with $N=10$. Both $T_1$ and $T_2$ range from 0.1 to 1.}
	\label{imperfect1}
\end{figure}

\begin{figure}[htbp]
	\centering
	\includegraphics[width=8cm]{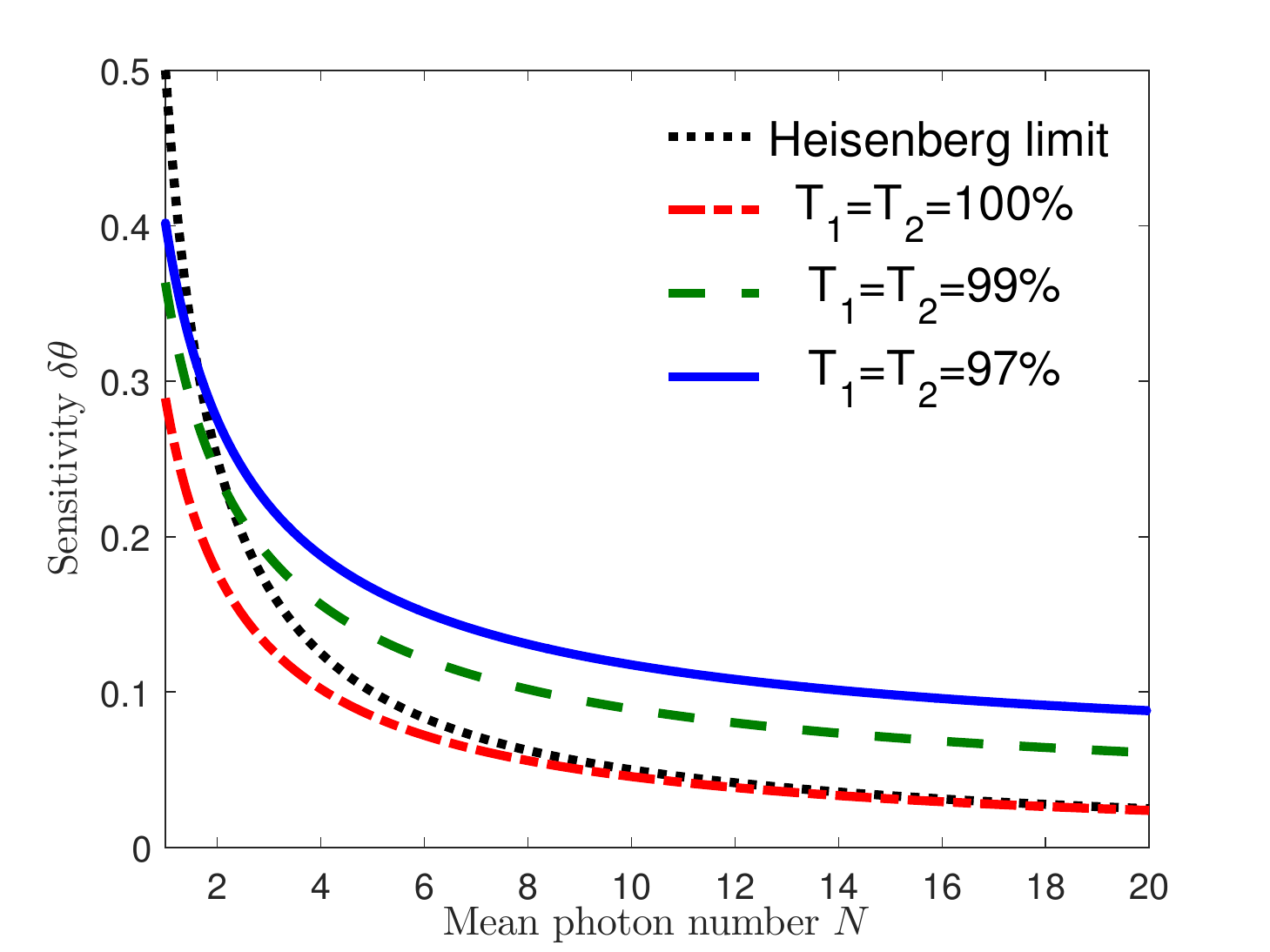}
	\caption{ Optimal sensitivity with parity detection as a functions of transmissivities, $T_1$ and $T_2$, where $N$ ranges from 1 to 20.}
	\label{loss1}
\end{figure}

To intuitively observe the impacts of imperfect state generation on the detection performance, in Fig. \ref{imperfect1} we plot the variances of signal visibility and sensitivity with different $T_1$ and $T_2$.
For a nearly ideal experimental environment (both $T_1$ and $T_2$ are greater than 0.9), one can see that the system can provide impressive visibility and sensitivity.
As to visibility, we can find that, same total generation loss ($T_1 +T_2 =$ Constant), the visibility of equal loss in two paths is always inferior to the unequal loss case, e.g., $V\left(T_1=T_2=0.5\right) <V\left( T_1=0.3, T_2=0.7\right) $.
For the large total generation loss (about $T_1 +T_2 \le 1$), we also find a similar situation that the sensitivity of equal loss in two paths is less than the unequal loss case.
However, once total generation loss leaves this region, the sensitivity of equal loss in two paths is slightly superior to that of unequal loss.
Additionally, in Fig. \ref{loss1} we plot sensitivities with different transmissivities, and we suppose that $T_1 = T_2$.
It can be seen from Fig. \ref{loss1} that the system's sensitivity will be gradually worse than HL with the increase of loss.
Meanwhile, the sensitivity difference between loss case and HL tends to a constant as increasing $N$.

\subsection{Imperfect output detection}
Next, we consider the second realistic factor: imperfect output detection.
The realistic detector usually has thermal noise and a detection efficiency which is less than 100\%.
We also insert a VBS, transmissivity $T$, in front of the ideal detector to simulate this factor \cite{PhysRevLett.108.233602}.
Unlike the VBS in imperfect generation, here, another input port of the VBS is a thermal state$\---$mean photon number $n_{\rm th}\---$rather than a vacuum to simultaneously model detection efficiency and thermal noise.
The explicit matrices of optical processes are also given in Appendix. Here we need to rewrite the transformation matrix $\bf S$ as $\mathbf S^{\rm R2} = {\mathbf S_{\rm{VBS}}^{\rm R2}}{\mathbf S_{\rm{QWP2}}^{\rm R2}}{\mathbf S_{\rm{RC}}^{\rm R2}}{\mathbf S_{\rm{QWP1}}^{\rm R2}}$.

According to the above description, we give the expectation value of parity operator,
\begin{equation}
{\left\langle {{{\hat \Pi }_B}} \right\rangle _{{\mathop{\rm R2}\nolimits} }} = \frac{1}{{\sqrt {{{\left[ {1 + 2{n_{\rm th}}\left( {1 - T} \right) + NT} \right]}^2} - {T^2}{{\sin }^2}\left( {2\theta } \right)N\left( {N + 2} \right)} }}. 
\end{equation}
Further, with the aid of classical Fisher information, the derivation of sensitivity is as below
\begin{equation}
\delta \theta  = \frac{{\sqrt {1 - {{\left\{ {{{\left[ {1 + 2{n_{\rm th}}\left( {1 - T} \right) + NT} \right]}^2} - {T^2}{{\sin }^2}\left( {2\theta } \right)N\left( {N + 2} \right)} \right\}}^{ - 1}}} }}{{\left| {\frac{{{T^2}\sin \left( {4\theta } \right)N\left( {N + 2} \right)}}{{{{\left\{ {{{\left[ {1 + 2{n_{\rm th}}\left( {1 - T} \right) + NT} \right]}^2} - {T^2}{{\sin }^2}\left( {2\theta } \right)N\left( {N + 2} \right)} \right\}}^{{3 \mathord{\left/{\vphantom {3 2}} \right.\kern-\nulldelimiterspace} 2}}}}}} \right|}}.
\end{equation}

\begin{figure}[htbp]
	\centering
	\includegraphics[width=8cm]{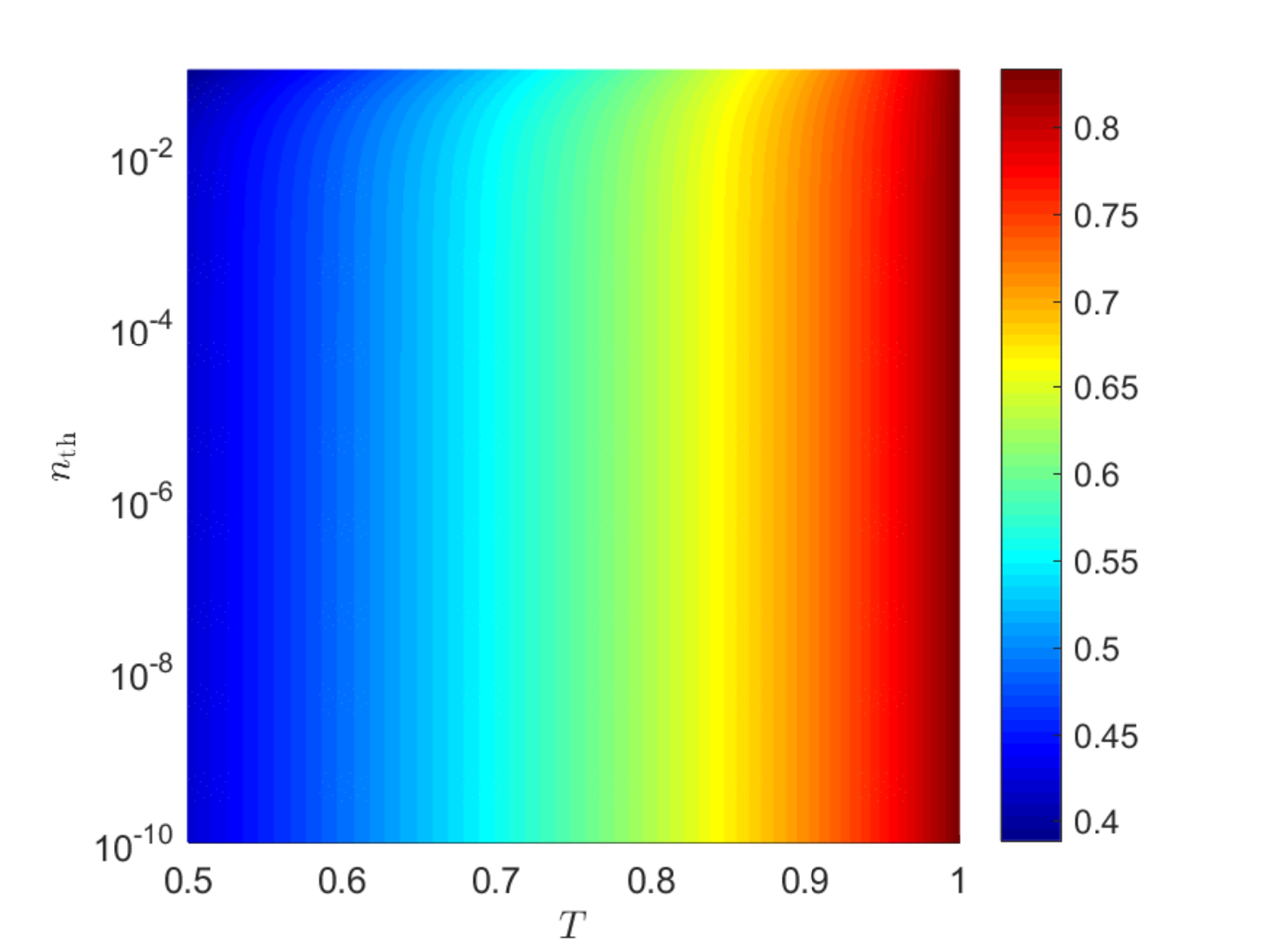}
	\includegraphics[width=8cm]{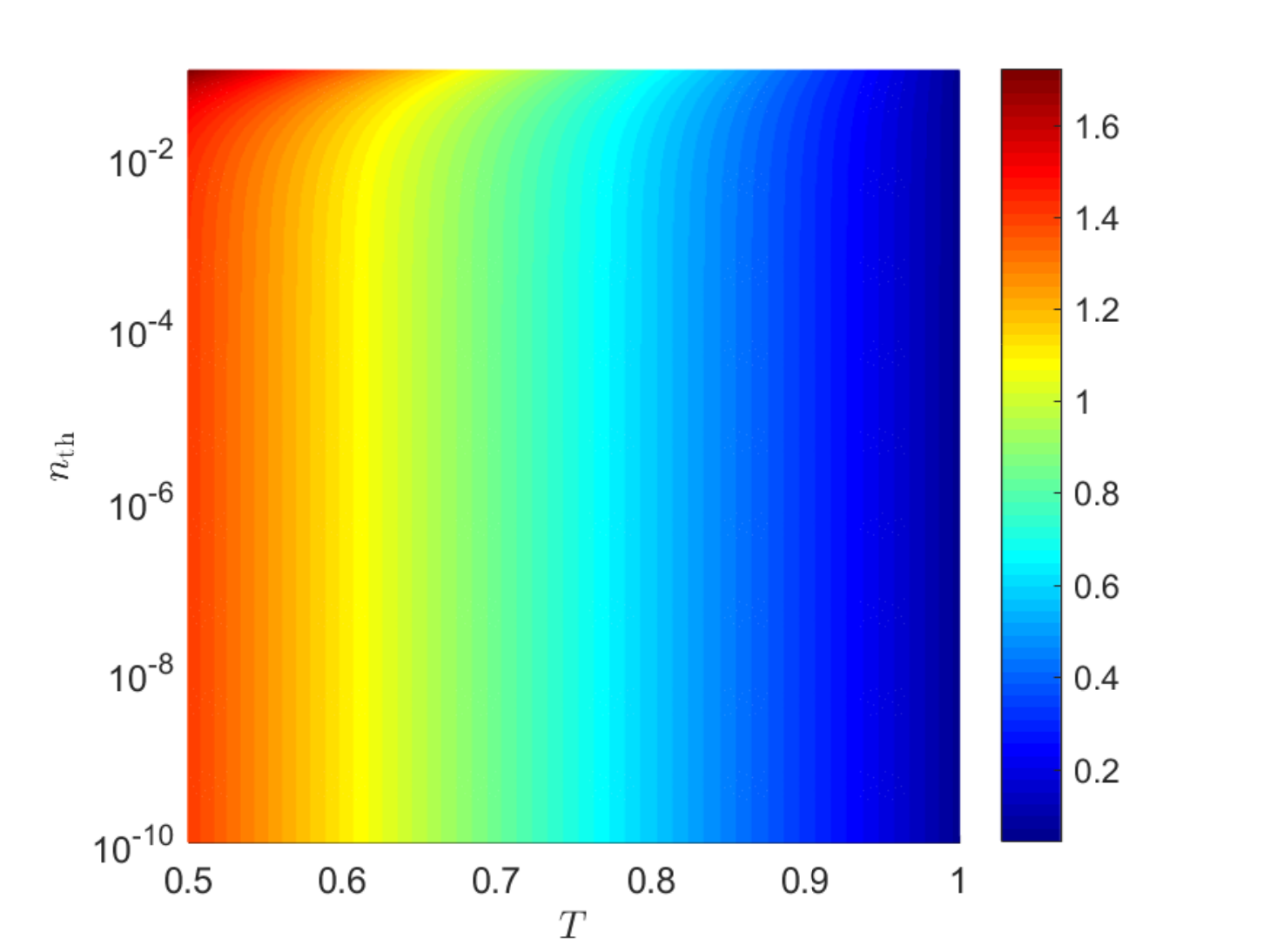}
	\caption{(a) Visibility with parity detection as a function of both detection efficiency $T$ and thermal photon number $n_{\rm th}$ in the case of $N=10$. $T$ ranges from 0.5 to 1, and $n_{\rm th}$ ranges from $10^{-10}$ to $10^{-1}$. (b) Optimal sensitivity with parity detection as a function of both detection efficiency $T$ and thermal photon number $n_{\rm th}$ in the case of $N=10$. $T$ ranges from 0.5 to 1, and $n_{\rm th}$ ranges from $10^{-10}$ to $10^{-1}$.}
	\label{imperfect2}
\end{figure}

In Fig. \ref{imperfect2}, we plot the impacts of thermal noise and detection efficiency on both the visibility and the sensitivity of output.
From Fig. \ref{imperfect2} we observe that the impacts of thermal noise on both visibility and sensitivity are almost negligible when the $n_{\rm th}$ is less than $10^{-2}$. 
For $n_{\rm th}\ge10^{-2}$, both visibility and sensitivity become worse with the increase of $n_{\rm th}$. 
In contrast, the variance of the sensitivity is more drastic than that of the visibility.
Additionally, both visibility and sensitivity uniformly increase as the growth of $T$.

\begin{figure}[htbp]
	\centering
	\includegraphics[width=8cm]{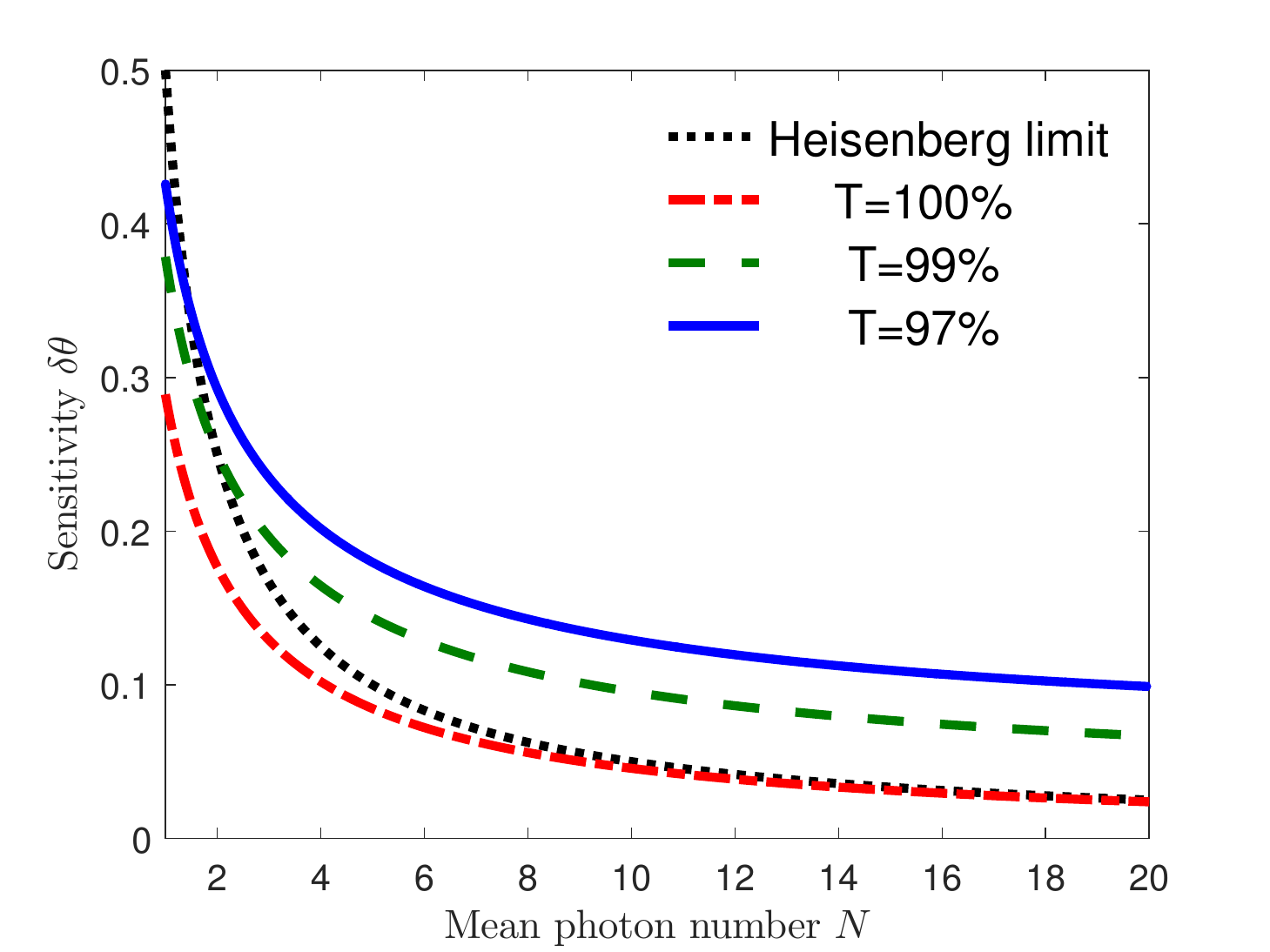}
	\caption{ Optimal sensitivity with parity detection as a function of detection efficiency $T$ in the case of $n_{\rm th}=0.1$, where $N$ ranges from 1 to 20.}
	\label{loss2}
\end{figure}

As a comparison of Fig. \ref{loss1}, we plot Fig. \ref{loss2} for sensitivities with different detection efficiencies.
The difference is that we let $n_{\rm th}=0.1$ here.
By contrast, we can see that the sensitivity in Fig. \ref{loss2} is slightly worse than that in Fig. \ref{loss1}, and this difference completely comes from the influence of thermal noise.
To show this, let us consider the conditions, $T_1 = T_2 = T$ and $n_{\rm th}=0$,
where $T_1$ and $T_2$ have defined in Eq. (17), $T$ and $n_{\rm th}$ have defined in Eq. (21).
An interesting phenomenon is that the sensitivity in Eq. (17) is equivalent to that in Eq. (21) for arbitrary rotating angle. 
This reveals that it is steer the same that linear photon loss occurs before or after RC under the situation of equal loss in two paths. 

\section{Conclusion} 
We report on a QPT scheme for rotating angle estimation using TMSV state and parity detection.
By means of the theory of symplectic matrix, we analyze the detection performances with lossless case.
A detection signal with both considerable visibility and sub-Heisenberg-limited sensitivity is obtained.
We also study the impacts of several realistic factor, like imperfect input generation and output detection.
For the visibility, the situation of unequal loss is always better than that of equal loss in the case of identical total loss.
As to the sensitivity, its change regulation is the same as that of visibility with about 50\% total loss. 
Once the total loss is lower than this limit, equal loss case can obtain better sensitivity compared to unequal one.
Additionally, the results indicate that the effects of thermal state on both visibility and sensitivity are fierce with $n_{\rm th} \ge 0.1$, whereas it can be ignored when the $n_{\rm th} \le 0.1$.
Finally, we find a novel phenomenon that there is no difference for linear photon loss before or after polarized rotation operation when the losses in two paths are equal.

\section*{Acknowledgments} 
This work is supported by the National Natural Science Foundation of China (Grant No. 61701139).


%

\section*{Appendix} 
\appendix

For imperfect input generation, the input mean and covariance matrix are given by 
\begin{equation}
{\bf{M}}_{\rm{in}}^ {\rm R1} = {\left( {\begin{array}{*{20}{c}}
		0 & 0 & 0 & 0& 0 & 0 & 0 & 0  \\
		\end{array}} \right)^\top},
\end{equation} 
\begin{equation}
\Gamma _{{\rm{in}}}^ {\rm R1}  = \Gamma _{{\rm{in}}} \oplus {{\rm I}\left( 4 \right)},
\end{equation}
where ${{\rm I}\left( 4 \right)}$ is a four-by-four identity matrix.
The optical element matrices of QWPs, RC and VBS are also provided with 

\begin{equation}
{\bf{S}}_{\rm QWP1}^ {\rm R1} = {\bf{S}}_{\rm QWP2}^ {\rm R1} = {\bf{S}}_{\rm QWP1} \oplus  {{\rm I}\left( 4 \right)},
\end{equation}

\begin{equation}
{\bf{S}}_{{\mathop{\rm RC}\nolimits} }^  {\rm R1}  = {\bf{S}}_{\rm RC} \oplus  {{\rm I}\left( 4 \right)},
\end{equation}

\begin{equation}
{{\mathbf{S}}^ {\rm R1}_{{\mathop{\rm VBS}\nolimits} }} = \left( {\begin{array}{*{20}{c}}
	{\sqrt {{T_1}} } & 0 & 0 & 0 & {\sqrt {1 - {T_1}} } & 0 & 0 & 0  \\
	0 & {\sqrt {{T_1}} } & 0 & 0 & 0 & {\sqrt {1 - {T_1}} } & 0 & 0  \\
	0 & 0 & {\sqrt {{T_2}} } & 0 & 0 & 0 & {\sqrt {1 - {T_2}} } & 0  \\
	0 & 0 & 0 & {\sqrt {{T_2}} } & 0 & 0 & 0 & {\sqrt {1 - {T_2}} }  \\
	{\sqrt {1 - {T_1}} } & 0 & 0 & 0 & { - \sqrt {{T_1}} } & 0 & 0 & 0  \\
	0 & {\sqrt {1 - {T_1}} } & 0 & 0 & 0 & { - \sqrt {{T_1}} } & 0 & 0  \\
	0 & 0 & {\sqrt {1 - {T_2}} } & 0 & 0 & 0 & { - \sqrt {{T_2}} } & 0  \\
	0 & 0 & 0 & {\sqrt {1 - {T_2}} } & 0 & 0 & 0 & { - \sqrt {{T_2}} }  \\
	\end{array}} \right).
\end{equation}

As to imperfect output detection, the input covariance and VBS matrices read
\begin{equation}
\Gamma _{\rm in}^ {\rm R2}  =  \Gamma_{\rm in} \oplus \left(2n_{\rm th} +1\right){\rm I}\left( 4 \right),
\end{equation}

\begin{equation}
{{\mathbf{S}}^ {\rm R2}_{{\mathop{\rm VBS}\nolimits} }} = \left( {\begin{array}{*{20}{c}}
	1 & 0 & 0 & 0 & 0 & 0 & 0 & 0  \\
	0 & 1 & 0 & 0 & 0 & 0 & 0 & 0  \\
	0 & 0 & {\sqrt {{T}} } & 0 & 0 & 0 & {\sqrt {1 - {T}} } & 0  \\
	0 & 0 & 0 & {\sqrt {{T}} } & 0 & 0 & 0 & {\sqrt {1 - {T}} }  \\
	0 & 0 & 0 & 0 &  - 1  & 0 & 0 & 0  \\
	0 & 0 & 0 & 0 & 0 &  - 1 & 0 & 0  \\
	0 & 0 & {\sqrt {1 - {T}} } & 0 & 0 & 0 & { - \sqrt {{T}} } & 0  \\
	0 & 0 & 0 & {\sqrt {1 - {T}} } & 0 & 0 & 0 & { - \sqrt {{T}} }  \\
	\end{array}} \right).
\end{equation}

Other matrices that are not mentioned are the same as the matrices of imperfect input generation.

\end{document}